\documentclass[aps,prb,preprint,showpacs,groupedaddress]{revtex4}
\usepackage{graphicx}
\usepackage{bm}
\usepackage{latexsym}
\usepackage{amssymb}

\begin{document}
\title{Thermodynamics of $S \geqslant 1$ ferromagnetic
Heisenberg chains with uniaxial single-ion anisotropy}

\author{I. Juh\'{a}sz Junger}
\author{D. Ihle}
\affiliation{Institut f\"{u}r Theoretische Physik,
Universit\"{a}t Leipzig, D-04109 Leipzig, Germany}
\author{J. Richter}
\affiliation{Institut f\"{u}r Theoretische Physik,
Otto-von-Guericke-Universit\"{a}t Magdeburg, D-39016 Magdeburg, Germany}

\date{\today}

\begin{abstract}
The thermodynamic properties of $S \geqslant 1$ ferromagnetic
chains with an easy-axis single-ion anisotropy are investigated at arbitrary
temperatures by both a Green-function approach, based on a decoupling of
three-spin operator products, and by exact diagonalizations of
chains with up to $N=12$ sites using periodic boundary conditions.
A good agreement between the results of both approaches is found.
For the $S=1$ chain, the temperature dependence of the specific heat reveals
two maxima, if the ratio of the anisotropy energy $D$ and the exchange energy
$J$ exceeds a characteristic value, $D/J >7.4$, and only one maximum for
$D/J <7.4$. This is in contrast to previous exact diagonalization data
for comparably small chains ($N \leqslant 7$)
using open boundary conditions. Comparing the theory with experiments on
di-bromo Ni complexes the fit to the specific heat yields concrete values
for $D$ and $J$ which are used to make predictions for the temperature
dependences of the spin-wave spectrum, the correlation length, and the
transverse magnetic susceptibility.
\end{abstract}


\maketitle

\section{INTRODUCTION}
In low-dimensional spin systems\cite{SRF04} the interplay of quantum
and thermal fluctuations and the effects of spin anisotropies on the
thermodynamics are of basic interest.
Whereas for Heisenberg antiferromagnets quantum
fluctuations occur already at $T=0$, in ferromagnets, possibly with an
easy-axis anisotropy, quantum fluctuations exist at nonzero temperatures
only. The study of systems with ferromagnetic exchange couplings, e.g.,
of the quasi-one-dimensional (1D) $S=1$ ferromagnet CsNiF$_3$ with
an easy-plane single-ion anisotropy,\cite{KV87} is also
motivated by the progress in the synthesis of new low-dimensional
materials, such as the frustrated $S=\frac{1}{2}$ quasi-1D
cuprates.\cite{Mat04} Likewise, the magnetic behavior of LaMnO$_3$,
a parent compound of the colossal magnetoresistance manganites,\cite{Ram97}
may be described by an effective spin $S=2$ model with a ferromagnetic
intraplane and an antiferromagnetic interplane
coupling, where neutron-scattering experiments\cite{Mou96} yield evidence
for a pronounced ferromagnetic short-range order (SRO) in the paramagnetic
phase and for an easy-axis single-ion anisotropy. To provide a good
analytical description of SRO and of the thermodynamics at arbitrary
temperatures, the standard spin-wave approaches cannot be adopted. Recently,
the Green-function equation of motion decoupling of second order and the
Green-function projection method with a two-operator basis,
respectively,\cite{KY72,
RS73,ST91,SSI94,WI97, ISF01, JIR04,SRI04} have been successfully applied
to quantum spin systems with anisotropies of different kind, where
most of the previous work was devoted to $S=\frac{1}{2}$ systems and only
in Refs.~\onlinecite{RS73} and \onlinecite{SSI94} low-dimensional isotropic
$S \geqslant 1$ ferromagnets have been considered. However, the study
of anisotropic $S \geqslant 1$ magnets is of current interest.\cite{GJL92,
Pot71,FJK00, FKS02} In Refs.~\onlinecite{Pot71,FJK00,FKS02} the
$S=1$ ferromagnet with an easy-axis single-ion anisotropy was investigated
in the random phase approximation (RPA) for the exchange term. By this
approach the paramagnetic phase and its SRO properties cannot be
described. Therefore, a second-order Green-function theory of SRO
for anisotropic $S \geqslant 1$ models, going one step beyond the RPA,
should be developed.

As a first step in this direction, we present a theory
for the $S \geqslant 1$ ferromagnetic chain with an easy-axis single-ion
anisotropy described by the model
\begin{equation}
H=-\frac{J}{2} \sum_{\langle i,j \rangle} \bm{S}_{i} \bm{S}_{j} -
D \sum_{i} (S_{i}^{z})^{2}
\label{ham}
\end{equation}
[$ \langle i,j \rangle $ denote nearest-neighbor (NN) sites] with
$J>0, \; D>0$, and $\bm{S}_{i}^{2}=S(S+1)$. The choice $D>0$ is motivated
by our emphasis on the temperature dependence of the specific heat,
which reveals a double maximum being more pronounced for $D>0$ than for
$D<0$,\cite{BLO75} and by the comparison with the experiments on di-bromo
Ni complexes\cite{KBD74} which may be described by the model (\ref{ham})
with $D>0$.

Furthermore, we perform exact finite-lattice diagonalizations
(ED) of $S=1$ chains with up to $N=12$ sites using periodic boundary
conditions which are critically analyzed in relation to the ED results
by Bl\"{o}te\cite{BLO75} for the specific heat using open boundary
conditions.

The paper is organized as follows. In Sec.~II the second-order
Green-function theory for the model (\ref{ham}) is developed extending
previous approaches for $S=\frac{1}{2}$ (Refs.~\onlinecite{KY72,ST91,
WI97,ISF01,JIR04,SRI04}) to $S \geqslant 1$ and rotation-invariant methods
for $S \geqslant 1$ (Refs.~\onlinecite{RS73} and \onlinecite{SSI94}) to
the case of an easy-axis on-site anisotropy. To test the theory, in
Sec.~III the limiting cases $J=0$ and $D=0$ are considered in comparison
with exact and ED results. The effects of spin anisotropy are explored
in Sec.~IV. The spin-wave spectrum, the spin susceptibility, and the
specific heat are investigated with the focus on the condition for the
existence of two maxima in the temperature dependence of the specific
heat. Moreover, the theory is compared with available experimental
data, and predictions for some relevant quantities are made. A summary
of our work can be found in Sec.~V.
\section{GREEN-FUNCTION THEORY}
The dynamic spin susceptibilities $\chi_{q}^{\nu \mu} (\omega) = - \langle
\langle S_{q}^{\nu};S_{-q}^{\mu} \rangle \rangle_{\omega} \; \; (\nu \mu =
+-,zz;\; -\pi \leqslant q \leqslant \pi)$, defined in terms of two-time
retarded commutator Green functions, \cite{EG79} are determined by the
projection method. \cite{WI97,ISF01,JIR04} Taking into account the breaking of
rotational symmetry by the single-ion spin anisotropy we choose, as in
Refs.~\onlinecite{ISF01} and \onlinecite{JIR04}, the two-operator basis
$\bm{A}^{\nu}=(S_{q}^{\nu}, i \dot S_{q}^{\nu})^{T} \; \; (\nu=+,z)$. Because
the model considered has up-down symmetry with respect to $S_{i}^{z} \rightarrow
-S_{i}^{z}$, we have $\langle S_{i}^{z} \rangle =0$. Neglecting the self-energy,
the matrix Green function
$ \langle \langle \bm{A}; \bm{A}^{+} \rangle
\rangle_{\omega} = [\omega - \bm{M}' \bm{M}^{-1}]^{-1} \bm{M}$
with the moment matrices
$ \bm{M} = \langle [\bm{A},\bm{A}^{+}] \rangle $
and
$ \bm{M}' = \langle [i \dot{\bm{A}} , \bm{A}^{+}] \rangle $
yields
\begin{equation}
\chi_{q}^{\nu \mu}(\omega)=-\frac{M_{q}^{\nu\mu}}{\omega^{2}-
(\omega_{q}^{\nu \mu})^{2}} ,\; \; \;
(\omega_{q}^{\nu \mu})^{2}=M_{q}^{(3)\nu\mu}/M_{q}^{\nu\mu},
\label{gf}
\end{equation}
where $M_{q}^{\nu\mu} = \langle [i \dot{S}_{q}^{\nu} , S_{-q}^{\mu}] \rangle$
and $M_{q}^{(3) \nu\mu} = \langle [- \ddot{S}_{q}^{\nu} , -i \dot{S}_{-q}^{\mu}]
\rangle$. The first spectral moments are given by the exact expressions
\begin{equation}
M_{q}^{+-}=4 J C_1 (1- \cos q) +2 D [3 C_0^{zz} - S(S+1)],
\label{mp}
\end{equation}
\begin{equation}
M_{q}^{zz}=2 J C_1^{+-} (1- \cos q).
\label{mz}
\end{equation}
The correlation functions $C_{n}=\frac{1}{2} C_{n}^{+-} + C_{n}^{zz} $ and $
C_{n}^{\nu \mu}= \langle S_0^{\nu} S_n^{\mu} \rangle = \frac{1}{N}
\sum_{q} C_{q}^{\nu \mu} \text{e}^{iqn}$
with $C_{q}^{\nu \mu}= \langle S_q^{\nu} S_{-q}^{\mu} \rangle $
are calculated by the spectral theorem, \cite{EG79} analogous to
Ref.~\onlinecite{JIR04}, as
\begin{equation}
C_{q}^{\nu \mu}=\frac{M_{q}^{\nu \mu}}{2 \omega_{q}^{\nu \mu}} [1+2
n(\omega_{q}^{\nu\mu})] + D_{q}^{\nu \mu} ,
\label{cq}
\end{equation}
\begin{equation}
D_{q}^{\nu \mu}= \lim_{\omega \to 0 } \frac{\omega}{2}
\langle \langle S_{q}^{\nu} ; S_{-q}^{\mu} \rangle \rangle_{\omega}^{(+)},
\label{dq}
\end{equation}
where $n(\omega)=( \text{e}^{\omega/T}-1)^{-1}$ and $\langle \langle \cdots ;
\cdots \rangle \rangle^{(+)}$ denotes the anticommutator Green function. The
on-site correlators $C_{0}^{\nu \mu}$ are related by the sum rule
\begin{equation}
C_{0}^{+-}+C_{0}^{zz}=S(S+1)
\label{sr}
\end{equation}
which follows from the operator identity
$\bm{S}_{i}^{2}=S_{i}^{+} S_{i}^{-} - S_{i}^{z} +(S_{i}^{z})^2 $
and $\langle S_{i}^{z} \rangle =0 $.

To obtain the spectra $\omega _{q}^{\nu \mu}$ in Eq.~(\ref{gf}) in terms of
two-spin correlation functions we approximate the time evolution of the spin
operators $ - \ddot{S}_{q}^{\nu}$ in the spirit of the schemes proposed in
Refs.~\onlinecite{KY72,RS73,ST91,SSI94,WI97,ISF01,JIR04}. That is, taking the
site representation the products of three spin operators in
$ - \ddot{S}_{i}^{\nu}$ are expressed in terms of one spin operator. Then, the
projection method neglecting the self-energy becomes equivalent to the equation
of motion decoupling in second order.

In $ - \ddot{S}_{i}^{+}$ we decouple the
operators along NN sequences $\langle i,j,l \rangle$ as \cite{JIR04}
\begin{equation}
S_{i}^{+} S_{j}^{+} S_{l}^{-} =
\alpha^{+-} \langle S_{j}^{+} S_{l}^{-} \rangle S_{i}^{+} +
\alpha^{+-} \langle S_{i}^{+} S_{l}^{-} \rangle S_{j}^{+}.
\label{entk1}
\end{equation}
Here, following the investigation of the isotropic ferromagnet,
\cite{ST91,SSI94} the dependence on the relative site positions
of the vertex parameters (cf.~Ref.~\onlinecite{WI97}) is neglected.

For $ S \geqslant 1$, in $ - \ddot{S}_{i}^{+}$ there appear products of three
spin operators with two coinciding sites which we decouple as proposed
in Refs.~\onlinecite{RS73} and \onlinecite{SSI94},
\begin{equation}
S_{i}^{+} S_{j}^{-} S_{j}^{+} =
\langle S_{j}^{-} S_{j}^{+} \rangle S_{i}^{+} +
\lambda^{+-} \langle S_{i}^{+} S_{j}^{-} \rangle S_{j}^{+}.
\label{entk2}
\end{equation}
Furthermore, for $D \neq 0, \; - \ddot{S}_{i}^{+}$ contains the term $D^2 A_i $
with%
\begin{equation}
A_{i} \equiv S_{i}^{+} (S_{i}^{z})^2 + 2 S_{i}^{z} S_{i}^{+} S_{i}^{z} +
(S_{i}^{z})^2 S_{i}^{+}.
\label{iii}
\end{equation}
For $ S=\frac{1}{2}$ we have $A_{i}=0$, and for $S=1$ we get $A_{i}=S_{i}^{+}$
(Ref.~\onlinecite{FKS02}) using the relation $(S_{i}^{z})^{2} S_{i}^{+} =
S_{i}^{z} S_{i}^{+}$ (Ref.~\onlinecite {JA00}).
To obtain a reasonable approximation of $A_{i}$ for $S>1$, we calculate
exactly the average $\langle A_{i} S_{i}^{-} \rangle (T)$ at $T=0$
 and $T \rightarrow \infty$. We obtain $\langle A_{i} S_{i}^{-} \rangle (0) =
(2 S-1)^{2} \langle S_{i}^{+} S_{i}^{-} \rangle (0) $ with
$\langle S_{i}^{+} S_{i}^{-} \rangle (0) = C_{0}^{+-}(0) =S $ and
$\lim_{T \to \infty} \langle A_{i} S_{i}^{-} \rangle = \frac{1}{5}
[4S (S+1)-3] \lim_{ T \to \infty} \langle S_{i}^{+} S_{i}^{-} \rangle$ with
$\lim_{ T \to \infty} \langle S_{i}^{+} S_{i}^{-} \rangle =\frac{2}{3} S (S+1)$.
Due to those results, for $S>1$ we approximately replace $A_{i}$
by
\begin{equation}A_{i}=\eta (2S-1)^{2} S_{i}^{+}
\label{nah}
\end{equation}
with $\eta(T=0)=1$ and $\eta(T \rightarrow \infty)=
\frac{4S(S+1)-3}{5(2S-1)^{2}}$. Note that Eq.~(\ref{nah}) holds exactly for
$S=1$ with $ \eta (T) =1$. Considering the ratio $R \equiv \eta/ C_{0}^{zz}$,
for$ S=2 \; (3)$ we have $ \lim_{T \to \infty} R = 0.23 \; (0.09)$ as compared
with $ R(0)=S^{-2} = 0.25 \; (0.11)$. Accordingly, for $ 1 < S \leqslant 3, \;
R(T)$ depends only weakly on temperature. Neglecting this dependence,
i.e., taking $R(T)=R(0), \; \eta(T)$ in Eq.~(\ref{nah}) may be calculated
in a reasonable approximation as
\begin{equation}
\eta (T) = \frac{1}{S^{2}} C_{0}^{zz} (T).
\label{etat}
\end{equation}
In $-\ddot{S}_{i}^{zz}$ we adopt the decouplings (cf. Refs.~\onlinecite{JIR04},
\onlinecite{ISF01}, and \onlinecite{SSI94})
\begin{equation}
S_{i}^{z} S_{j}^{+} S_{l}^{-} = \alpha^{zz} \langle S_{j}^{+} S_{l}^{-} \rangle
S_{i}^{z},
\label{entk3}
\end{equation}
\begin{equation}
S_{i}^{-} S_{j}^{z} S_{j}^{+} = \lambda^{zz} \langle S_{j}^{+} S_{i}^{-} \rangle
S_{j}^{z}.
\label{entk4}
\end{equation}
Finally, we obtain the spectra
\begin{equation}
(\omega_{q}^{+-})^2 = (1-\cos q) \{
\Delta^{+-} + 4 J^{2} \alpha^{+-}C_1 (1-
\cos q) \} + (\omega_{0}^{+-})^2,
\label{op}
\end{equation}
\begin{eqnarray}
\Delta^{+-} &=& J^{2} \{S (S+1)+ C_{0}^{zz} + 2
\lambda^{+-} C_1+ 2 \alpha^{+-}(C_2-3 C_1)\} \nonumber
\\&+& 2 D J \{ 2 \lambda^{+-} C_{1}^{zz} + 3 C_{0}^{zz}
-S (S+1) \},
\label{dp}
\end{eqnarray}
\begin{equation}
(\omega_{0}^{+-})^2 = 2 D J \{ S (S+1) -3 C_{0}^{zz}+
\lambda^{+-}(2 C_{1}^{zz}-C_{1}^{+-})\}+ \eta (2 S-1)^2 D^2,
\label{o0}
\end{equation}
\begin{equation}
(\omega_{q}^{zz})^2=(1-\cos q) \{ \Delta^{zz}+ 4 J^{2} \alpha^{zz}
C_1^{+-}(1-\cos q) \},
\label{oz}
\end{equation}
\begin{equation}
\Delta^{zz}= 2 J^{2} \{S (S+1)-C_0^{zz}+ \alpha^{zz} (C_2^{+-}-3
C_1^{+-}) \} + 2 J (J-2D)\lambda^{zz}C_1^{+-}.
\label{dz}
\end{equation}
To calculate the correlation functions $C_{n}^{\nu \mu}$ from Eq.~(\ref{cq}),
in particular the term $D_{q}^{\nu \mu}$ given in Eq.~(\ref{dq}),
we follow the reasonings of our previous paper.\cite{JIR04} We obtain
$D_{q}^{+-}=0$, because $\omega_{q = 0}^{+-} \neq 0$ for $D \neq 0$, and
$D_{q}^{zz}=\sum_{n}C_{n}^{zz} \delta_{q,0}$. Then, the longitudinal correlation
functions are calculated as
\begin{equation}
C_{n}^{zz}=\frac{1}{N} \sum_{q (\neq 0)} C_{q}^{zz} \text{e}^{i q n}+ C^{zz}
\label{cnz}
\end{equation}
with $ C^{zz}=\frac{1}{N} \sum_{n} C_{n}^{zz} $ and $C_{q}^{zz}$ given by the
first term in Eq.~(\ref{cq}). Note that the term $C^{zz}$ describes long-range
order in the infinite system. In
previous work \cite{ST91,WI97,ISF01,SRI04} such
terms are introduced by hand and interpreted as condensation parts.

By Eqs.~(\ref{gf}),(\ref{mz}),(\ref{oz}), and (\ref{dz}) we get the longitudinal
static susceptibility
\begin{equation}
\chi_{q}^{zz} = \chi_{0}^{zz} \{ 1+ 4 \alpha^{zz} C_{1}^{+-}
(\Delta^{zz})^{-1} (1- \cos q) \}^{-1} ,
\label{hiqz}
\end{equation}
where
$
\chi_{0}^{zz} = \frac{2 C_{1}^{+-}}{J \Delta^{zz}}.
$
Expanding the denominator  for small $q$ up to $O(q^2)$ we obtain the
correlation length $ \xi^{zz}= \sqrt{2 \alpha^{zz} C_{1}^{+-}/ \Delta^{zz}}$.

The uniform static susceptibilities $ \chi_{0}^{\nu \mu}$ may be also expressed
in terms of $ C_{n}^{\nu \mu} $. From Eqs.~(\ref{gf}) and (\ref{cq}) with
$\lim_{q \to 0} C_{q}^{zz} = T \lim_{q \to 0} \chi_{q}^{zz}$, we get
\begin{equation}
\chi_{0}^{zz} = \frac{1}{T} \sum_{n} C_{n}^{zz},
\label{hi0z}
\end{equation}
which agrees with the general formula of thermodynamics in the case $\langle
S_{i}^{z} \rangle = 0 $. For $\chi_{0}^{+-}$ we obtain
\begin{equation}
\chi_{0}^{+-} = g \sum_{n} C_{n}^{+-}
\label{hi0p}
\end{equation}
 with $g=2 \{ \omega_{0}^{+-}
[1+2 n(\omega_{0}^{+-})]\}^{-1}$, following from Eqs.~(\ref{gf}), (\ref{mp}),
(\ref{cq}), and (\ref{op}).%

For large temperatures, $T \gg \omega_{q}^{+-}$, the static susceptibilities and
the structure factors $ C_{q}^{\nu \mu }$ are related by
\begin{equation}
\chi_{q}^{\nu \mu}= \frac{1}{T} C_{q}^{\nu \mu} =\frac{1}{T} \sum_{n}
C_{n}^{\nu \mu} \text{e}^{-iqn}.
\label{hiq}
\end{equation}
At very high temperatures, in Eq.~(\ref{hiq}) only the $n=0$ term may be taken
into account, and, with $C_{0}^{+-} = 2 C_{0}^{zz} = \frac{2}{3} S (S+1)$,
we get the Curie law $\chi_{q}^{+-} = 2 \chi_{q}^{zz} = 2 S(S+1)/3T$.

To provide a better comparison of the Green-function theory with ED data, it is
useful to consider the theory also for finite systems with periodic boundary
conditions. For a ring with an even number N of spins we have the discrete $q$
values $q_{i}=\frac{2 \pi}{N} n_{i} $ with $-\frac{N}{2} \leqslant n_{i}
\leqslant \frac{N}{2}-1$ and $\frac{N}{2}+1$ correlators $C_{n}^{\nu \mu}$ with
$ 0 \leqslant n \leqslant \frac{N}{2}$. In the calculation of $C_{n}^{zz}$
according to Eq.~(\ref{cnz}) we must take care of the term $C^{zz}$ which, for
finite $N$, is finite at arbitrary temperatures and is given by
\begin{equation}
C^{zz}=\frac{1}{N}\left(C_{0}^{zz}+C_{N/2}^{zz}+ 2 \sum_{n=1}^{N/2-1}
C_{n}^{zz} \right).\label{konst}
\end{equation}
Then, it turns out that two equations for $C_{n}^{zz}$ are linearly dependent.
As additional equation, we use the expression of $\chi_{0}^{zz}$ in
Eq.~(\ref{hiqz}) in terms of $C_{n}^{zz}$ according to Eq.~(\ref{hi0z}), i.e.
\begin{equation}
\frac{2C_{1}^{+-}}{J \Delta^{zz}}=\frac{1}{T} \sum_{n} C_{n}^{zz}.
\label{gl}
\end{equation}
\section{LIMITING CASES}
To test the approximations made for $S \geqslant 1$ in addition to those for
$S=\frac{1}{2}$, in particular the decouplings (\ref{entk2}) and (\ref{entk4}),
where $\lambda^{\nu \mu} (S=\frac{1}{2})=0$, and the replacement (\ref{nah})
with Eq.~(\ref{etat}), we first consider the limiting cases $J=0$ and $D=0$.

In the $J=0$ limit, by Eqs.~(\ref{cq}), (\ref{mp}), (\ref{sr}), and (\ref{o0}) we
obtain
\begin{equation}
C_{0}^{+-}=\frac{2 S (S+1)}{3 + \sqrt{\eta} (2 S-1) [1+2
n(\omega_{0}^{+-})]^{-1}} ,
\label{c0}
\end{equation}
\begin{figure}
\centering
\includegraphics{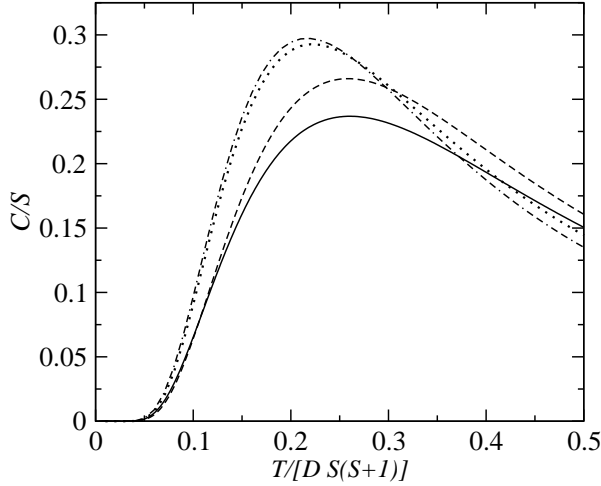}
\caption{ $J=0$ limit: Specific heat
for $S>1$. The Green-function theory for $S=\frac{3}{2}$ (dashed) and $S=2$
(solid) is compared with the exact results for $S=\frac{3}{2}$ (dotted)
and $S=2$ (dot-dashed), respectively.}
\label{fig1}
\end{figure}
where $\omega_{0}^{+-} = \sqrt{\eta} (2 S-1) D$ and $\eta$ is calculated by
Eq.~(\ref{etat}). The longitudinal on-site correlator
$C_{0}^{zz}$ is obtained from
Eq.~(\ref{sr}). At $T=0$ and $ T \rightarrow \infty$ we get $C_{0}^{zz} =S^2$
and $C_{0}^{zz} =\frac{1}{3} S(S+1)$, respectively, agreeing with the exact
results. Figure \ref{fig1} shows the specific heat for $S=\frac{3}{2}$ and
$S=2$ derived from $C_{0}^{zz}$, where
the temperatures of the maximum $T_{m}^{C} (S=\frac{3}{2})=0.97 D$ and
$T_{m}^{C} (S=2) =1.6 D$ nearly agree with the exact values
$T_{m}^{C}=0.83 D$ and $T_{m}^{C}=1.4 D$ for $S=\frac{3}{2}$ and
$S=2$, respectively.
This yields a justification for the approximations (\ref{nah}) and (\ref{etat}).
The result for $S=1$, not depicted in Fig.~\ref{fig1} for clarity, shows
qualitatively the same temperature dependence as that for $S>1$; it is
exact, because Eq.~(\ref{nah}) becomes the exact relation $A_{i}=S_{i}^{+}$.

In the
$D=0$ limit, we have $C_{n}^{+-}=2 C_{n}^{zz}, \; \alpha^{\nu \mu} =\alpha, \;
\lambda^{\nu \mu} =\lambda, \; \Delta^{\nu \mu} = \Delta $, and $\omega_{q}^{\nu
\mu}=\omega_{q}$. The vertex parameter $\alpha (T)$ is determined by
Eq.~(\ref{sr}), $C_{0}^{zz}=S(S+1)/3$. To derive an equation for $\lambda(T)$,
we first consider the long-range ordered ground state with $\xi^{-1}(0)=0$
corresponding, by Eq.~(\ref{hiqz}), to $\Delta(0)=0$. Then, by Eq.~(\ref{op})
we have $\omega_{q}=2J\sqrt{2 \alpha C_{1}^{zz}} (1- \cos q)$ and,
by Eq.~(\ref{cnz}),
$C_n^{zz}=\sqrt{\frac{C_{1}^{zz}}{2 \alpha}} \delta_{n,0}+C^{zz}$. Taking into
account the exact result $C_{n \neq 0}^{zz}(0)=\frac{1}{3}S^2$ we get $\alpha
(0) = \frac{3}{2}$ and $\lambda (0) = 2-\frac{1}{S}$
(cf.~Ref.~\onlinecite{SSI94}). At non-zero temperatures there is no long-range
order, i.e. $C^{zz}=0$ and $\Delta > 0$. To improve the approximation of
Ref.~\onlinecite{SSI94}, $\lambda(T)=\lambda(0)$, we first derive the exact
high-temperature series expansion of $C_{1}^{zz}$ up to $O(T^{-2})$,
\begin{equation}
C_{1,ex}^{zz}=\left [\frac{S(S+1)}{3}\right ]^{2} \left ( \frac{J}{T}
- \frac{1}{4} \frac{J^2}{T^2} \right ) + O(T^{-3}).
\label{c1z_se}
\end{equation}
Expanding Eq.~(\ref{cnz}) for $n=1$ and $n=0$ up to $O(T^{-1})$ and
using, for $n=0$, Eq.~(\ref{c1z_se}) we obtain
\begin{equation}
C_{1}^{zz}=\left[ \frac{S (S+1)}{3} \right]^2 \alpha_{0} \frac{J}{T},
\label{c1}
\end{equation}
\begin{equation}
C_{0}^{zz}=\frac{S (S+1)}{3} \left\{ 1 - \frac{1}{12} [3+ 4 S(S+1)
(\lambda_{0} - \alpha_{0})] \frac{J}{T} \right \},
\label{c0z}
\end{equation}
where $\alpha_{0}$ and $\lambda_{0}$ are the lowest orders in the expansions of
$\alpha (T)$ and $\lambda(T)$, respectively. The comparison with the exact
results Eq.~(\ref{c1z_se}) and $C_{0}^{zz}= S (S+1)/3$ yields
\begin{equation}
\alpha_{0}=1, \; \; \lambda_{0}=1-3 [4 S(S+1)]^{-1}.
\label{parint}
\end{equation}
The result $\alpha_{0}=1$ confirms the general suggestion (cf.
Refs.~\onlinecite{ST91}, \onlinecite{WI97}, \onlinecite{ISF01}) that the
vertex parameters $\alpha$ approach unity at high temperatures. Considering
the ratio $Q \equiv \lambda / \alpha$, for $S=1, 2$ and 3 we have $\lim_{ T \to
\infty} Q =0.63,\; 0.88$, and 0.94, respectively, as compared with $Q(0)=0.67,1
$, and 1.1. Accordingly, for $1 \leqslant S \leqslant 3, \, Q(T)$ is only weakly
temperature dependent. Setting $Q(T)=Q(0), \; \lambda (T)$ may be calculated
in a rather good approximation as
\begin{equation}
\lambda (T) = \frac{2}{3} \left ( 2-\frac{1}{S} \right) \alpha (T).
\label{lt}
\end{equation}
\begin{figure}
\centering
\includegraphics{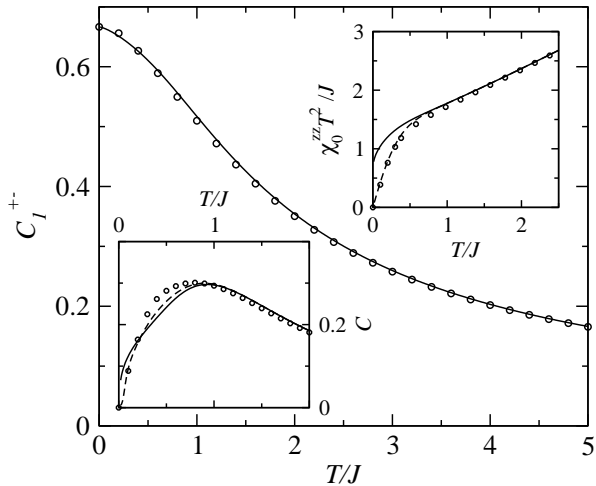}
\caption{$D=0$ limit: Nearest-neighbor correlation function $C_{1}^{+-}$,
longitudinal spin susceptibility $\chi_{0}^{zz}$
(upper inset), and specific heat $C$ (lower inset) for $S=1$, as
obtained by the Green-function theory in the thermodynamic limit (solid) and for
a finite system with $N=12$ (dashed) in comparison with the ED data
for $N=12$ ($\circ$). The Green-function results for $C_{1}^{+-}$ obtained for
$N=12$ and $N \to \infty$ agree within the accuracy of drawing.}
\label{fig2}
\end{figure}
In Fig.~\ref{fig2} our results for $S=1$ are plotted, where a remarkably good
agreement of the Green-function theory for $N=12$ with the ED data is found.
This justifies the decouplings (\ref{entk2}) and (\ref{entk4}) with $\lambda$
calculated by Eq.~(\ref{lt}). Considering the specific heat, the temperature of
the maximum in the theory, $T_{m}^{C}=0.9 J$ for both $N=12$ and $N \to \infty$,
only slightly deviates from the ED result $T_{m}^{C}=0.8J$. Note that $T_{m}^{C}$
in the semiclassical approach of Ref.~\onlinecite{CTV00} agrees with the ED
value. Concerning the uniform static susceptibility in the thermodynamic limit,
we have (see also Ref.~\onlinecite{SSI94}) $\lim_{T \to 0} \chi_{0}^{zz}
T^{2}/J = \frac{2}{3} S^{4}$.
This low-temperature behavior, $\chi_{0}^{zz} \propto T^{-2}$, qualitatively
agrees with the result of the renormalization-group approach of
Ref.~\onlinecite{Kop89}, but quantitatively deviates from the finding
$\lim_{T \to 0} \chi_{0}^{zz} T^{2}/J=1.58 S^{2}$. \cite{Kop89}

Finally, let us compare our theory with the Green-function approach of
Ref.~\onlinecite{BZS96} for the $S=1$ antiferromagnetic Heisenberg chain. There,
instead of the decoupling (\ref{entk4}), the left-hand side is rewritten as
$S_{i}^{-}S_{j}^{z} S_{j}^{+} = S_{i}^{-} S_{j}^{+} S_{j}^{z} + S_{i}^{-}
S_{j}^{+} $. The first term is decoupled analogous to Eq.~(\ref{entk4}) as
$\alpha\langle S_{i}^{-} S_{j}^{+} \rangle S_{j}^{z}$, whereas the second term
yields a contribution to $i \dot{S}_{i}^{z}$. This results in a gap $\Delta$ in
$\omega_{q}^{zz}$ at $q=0$ which, for $D=0$, is given by $\Delta = J$. In
Ref.~\onlinecite{BZS96}, $\Delta$ is interpreted as a Haldane gap. As we have
verified, $\Delta$ is independent of $S$. However, for $S=\frac{3}{2}$, for
example, there is no Haldane gap. That means, the gap $\Delta$ is an artefact of
the approach of Ref.~\onlinecite{BZS96} employing commutation before decoupling.
According to our experience (see, e.g., the Green-function theory for the $t-J$
model\cite{WI98}) such a procedure should be avoided. Furthermore, we argue that
the approach of Ref.~\onlinecite{BZS96} yields $\Delta \neq 0$ for the $S=1$
antiferromagnet also in higher dimensions and for the $S \geqslant 1$
ferromagnetic chain. Concluding, contrary to the reasonings of
Ref.~{\onlinecite{BZS96}}, the Haldane physics cannot be captured by the
second-order Green-function theory.%
\section{EFFECTS OF SPIN ANISOTROPY}
To complete our Green-function scheme for the model (\ref{ham}) with $D>0$ and
$J>0$ (hereafter, we set $J=1$), the four parameters $\alpha^{\nu \mu} $ and
$\lambda^{\nu \mu}$ have to be determined. In the ground state, for $D>0$ we
have the exact results
\begin{equation}
C_{n}^{+-} (0)= S \delta_{n,0} ; \; \; \; C_{n}^{zz} (0)= S^{2}
\label{ex}
\end{equation}
so that $C_{n \neq 0} (0)= S^{2}$. By Eq.~(\ref{cq}) we get $C_{n}^{+-} (0)=
\frac{1}{N} \sum_{q} \frac{M_{q}^{+-}}{2 \omega_{q}^{+-}} \text{e}^{i q n}$ and,
comparing with Eq.~(\ref{ex}),
\begin{equation}
M_{q}^{+-}=2 S \omega_{q}^{+-}.
\label{mq}
\end{equation}
Inserting $M_{q}^{+-}$ and $\omega_{q}^{+-}$ given by Eqs.~(\ref{mp}) and
(\ref{op}) to (\ref{o0}) with $\eta (0)=1$ [see Eq.~(\ref{nah})] and comparing
the coefficients in Eq.~(\ref{mq}) in front of $ (1- \cos q)^{n} \; (n=2
\text{ and } 0 \text{ or } 1)$, we obtain
\begin{equation}
\alpha^{+-}(0)=1; \; \; \lambda^{+-}(0)=1-\frac{1}{2S}.
\label{par0}
\end{equation}
\begin{figure}
\centering
\includegraphics{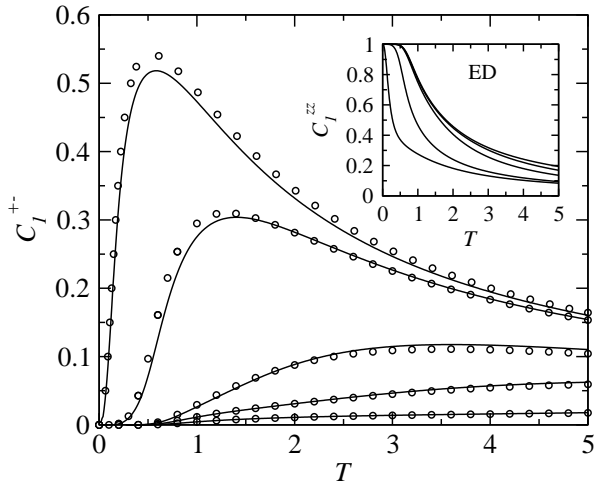}
\caption{Nearest-neighbor transverse spin correlation functions $C_{1}^{+-}$
for $S=1$ at $D=0.1,1,5,10$, and 25, from top to bottom, showing the
Green-function (solid) and ED results ($\circ, \; N=12$). The inset shows
the ED data for the nearest-neighbor longitudinal correlation functions
$C_{1}^{zz}$ at the same values of D with D increasing from bottom to
top, which are used as input to the Green-function theory.}
\label{fig3}
\end{figure}
Considering finite temperatures and suggesting $\lim_{T \to \infty} \alpha^{+-}
(T) = 1 $ (see Sec. III), we put $\alpha^{+-} (T)=1 $ because of $\alpha^{+-}
(0)=1$. Following the reasonings in the $D=0$ limit, for the ratio $Q^{+-}
\equiv \lambda^{+-}/ \alpha^{+-}$ we assume $Q^{+-} (T) = Q^{+-} (0)$, i.e.
$\lambda^{+-} (T)= \lambda^{+-} (0)$. The parameter $\alpha^{zz} (T) $ is
calculated from the sum rule (\ref{sr}). Concerning the remaining parameter
$\lambda^{zz}$ and the ratio $Q^{zz} \equiv \lambda^{zz}/ \alpha^{zz}$,
it turns out that $Q^{zz}$ has very different values in the
$T \rightarrow 0$ and $T \rightarrow \infty$ limits. Therefore,
we adjust $\lambda^{zz} (T) $ to the ED data for $C_{1}^{zz}
(T)$ which are depicted, for $S=1$, in the inset of Fig.~\ref{fig3}. Thus, we
have a closed system of equations for seven quantities ($C_{0}^{\nu \mu},
C_{1}^{+-},C_{2}^{\nu \mu}, \alpha^{zz}, \lambda^{zz}$) to be determined
self-consistently as functions of temperature.

As a first test of our approach, in Fig.~\ref{fig3} the NN correlation function
$C_{1}^{+-}$ for $S=1$ is plotted, where a very good agreement with the ED
results
is found. The correlator $C_{0}^{+-}$ (not shown) also agrees very well with the
ED data. As can be seen, we have $C_{1}^{+-} < 2 C_{1}^{zz} $; that is, due to
the easy-axis anisotropy the transverse correlations are suppressed as
compared with the  longitudinal correlations. The maximum in the temperature
dependence of $C_{1}^{+-}$ indicates the crossover from Ising-like to
Heisenberg-like behavior, where the maximum position increases with increasing
$D$.
\subsection{Spin waves}
At $T=0$, by Eq.~(\ref{mq}) with Eqs.~(\ref{mp}) and (\ref{ex}) we obtain the
spin-wave spectrum
\begin{equation}
\omega_{q}^{+-} (0) = 2S (1-\cos q)+ (2 S-1)D
\label{om0}
\end{equation}
with the spin-wave gap $\omega_{0}^{+-} (0) = (2S-1) D $. Let us point out that
the dispersion (\ref{om0}) agrees with the result obtained by the RPA and
the Anderson-Callen decoupling (see, e.g.,
Ref.~\onlinecite{FJK00}) given by  $B_{i} \equiv S_{i}^{+} S_{i}^{z} + S_{i}^{z}
S_{i}^{+} = 2\langle S^{z} \rangle \left \{ 1-\frac{1}{2S^2} \cdot [S(S+1) -
C_{0}^{zz}]\right \} S_{i}^{+}$; putting, at $T=0$, $\langle S^{z}\rangle = S $
and $ C_{0}^{zz}=S^{2}$ so that $B_{i}= 2 S -1$, the spectrum (\ref{om0})
results. In the RPA approach of Ref.~\onlinecite{FKS02}, where the $D$ term for
$S=1$ is treated exactly, we calculate $\chi_{q}^{+-} (\omega) = - 2 (\omega-
\omega_{q})^{-1}$ (correcting a misprint in Eq.~(51) of Ref.~\onlinecite{FKS02})
with $\omega_{q}$ given by Eq.~(\ref{om0}) with $S=1$.

Let us compare Eq.~(\ref{om0}) with previous spin-wave theories. The generalized
spin-wave theory by Becker \cite{Bec72} for $S=1$, which extends the
Holstein-Primakoff transformation to two sets of Bose operators treating the
single-ion anisotropy exactly,
yields $\omega_{0}^{+-} (0)=D$, in agreement with Eq.~(\ref{om0}). Contrary,
in the ordinary spin-wave theory (with only one Bose operator $a_{i}$),
$(S_{i}^{z})^2$ with $S_{i}^{z}=S-n_{i}$ and $n_{i}=a_{i}^{+} a_{i}$ is
approximated as $(S_{i}^{z})^{2}=S^2 - 2 S n_i$ neglecting the $n_i^2$ term.
This yields the wrong result $\omega_0^{+-}=2 S D$ violating the condition
$\omega_0^{+-}(S=\frac{1}{2})=0$. Note that such an approach was used to fit the
inelastic neutron-scattering data on LaMnO$_3$ on the basis of an effective
spin model with easy-axis single-ion anisotropy. \cite{Mou96} From our
results we conclude that this fit should be reconsidered by means of an
improved theory.

\begin{figure}
\centering
\includegraphics{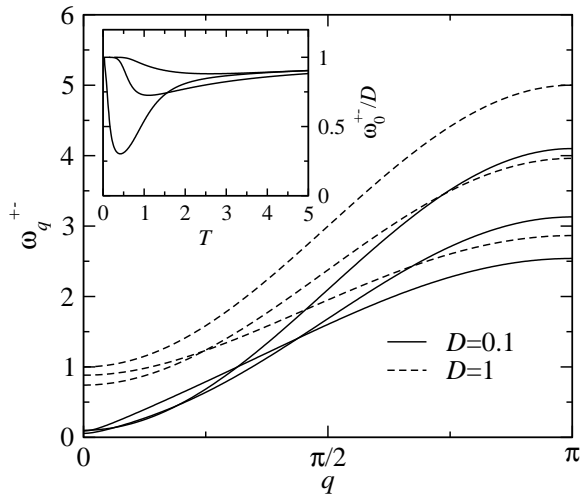}
\caption{Spin-wave spectrum $\omega_{q}^{+-}$ for $S=1$ at $T=0,1,$ and 5,
at $q=\pi$ from top to bottom. The inset shows the spin-wave gap
$\omega_{0}^{+-}$ at $D=0.1,1,$ and 5, at $T=1$ from
bottom to top.}
\label{fig4}
\end{figure}
\begin{figure}
\centering
\includegraphics{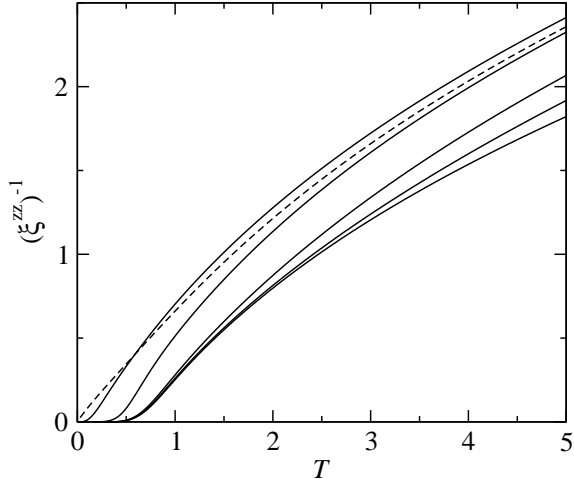}
\caption{Inverse correlation length for $S=1$ at $D=0.1,1,5,10$, and 25,
from top to bottom (solid), compared with the $D=0$ limit (dashed).}
\label{fig5}
\end{figure}
\begin{figure}
\centering
\includegraphics{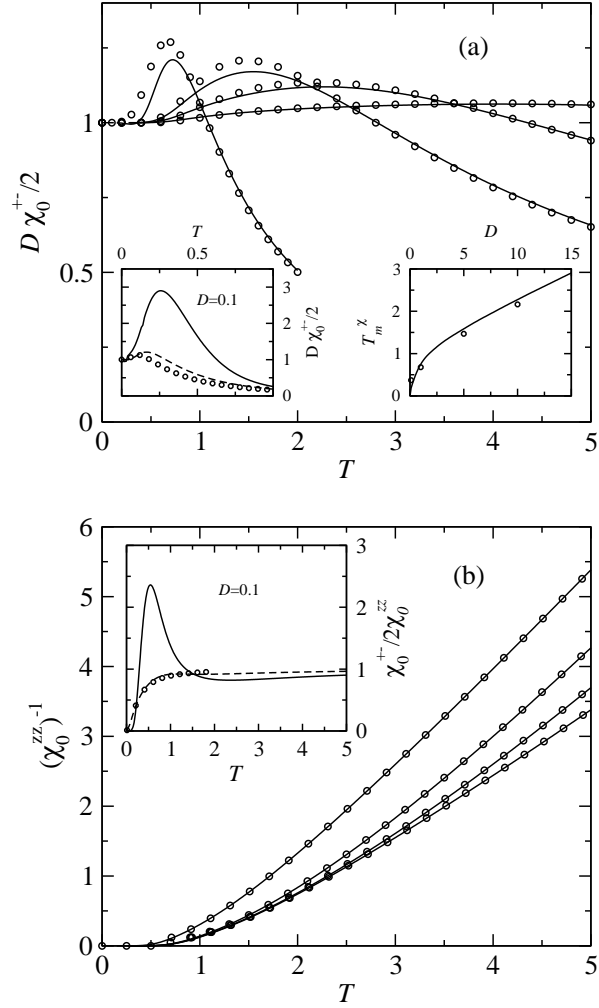}
\caption{Transverse (a) and longitudinal (b) uniform static spin
susceptibility for $S=1$ at $D=1,5,10$, and 25, from left to right,
obtained by the ED ($\circ$) for $N=8$ (a) and $N=12$ (b) and by the
Green-function theory (solid). In the insets on the left-hand side,
for $D=0.1$ the Green-function results for $N \rightarrow\infty $
(solid) and $N=8$ (dashed) are compared with the ED data
for $N=8$. In the inset on the right-hand side of (a) the position of the
susceptibility maximum $T_{m}^{\chi}$ vs $D$ is depicted.}
\label{fig6}
\end{figure}
In Fig.~\ref{fig4} the temperature dependence of the spin-wave spectrum for
$S=1$ is shown, where a spin correlation-induced flattening of the shape with
increasing temperature is observed. The spin-wave gap as function of temperature
exhibits a minimum and approaches the high-temperature limit $\lim_{T \to
\infty} \omega_{0}^{+-} (T) = \omega_{0}^{+-}(0)$. In the paraphase
($T>0$) with SRO, well-defined spin waves exist, if their wavelength is
much smaller than the correlation length, i.e., if $q \gg (\xi^{zz})^{-1}$.
To estimate the validity region of the spin-wave picture, in
Fig.~\ref{fig5} the inverse correlation length is plotted. For $D=0$ we get
$\lim_{T \to 0} \xi^{zz} T = S^2$ (cf. Ref.~\onlinecite{SSI94}) which nearly
agrees with the result of the renormalization-group approach, \cite{Kop89}
$\lim_{T\to 0} \xi^{zz} T = 1.14 S^2$. For $D>0$ the low-temperature behavior of
$\xi^{zz}$ is quite different. By Eq.~(\ref{hiqz}) we have $\xi^{zz}=\sqrt{2
\bar{\alpha}^{zz} / \Delta^{zz}}$ and $ (\xi^{zz})^{-2} \chi_{0}^{zz} =
C_{1}^{+-} / \bar{\alpha}^{zz}$ with $\bar{\alpha}^{zz} \equiv
\alpha^{zz}C_{1}^{+-}$, where the numerical evaluation yields a finite value
of $\bar{\alpha}^{zz}$ as $T \to 0$. Because $ (\chi_{0}^{zz})^{-1} (0)=0$,
$(\xi^{zz})^{-2}$ approaches zero as $T \to 0$ much stronger than $C_{1}^{+-}$
and $(\chi_{0}^{zz})^{-1}$ (compare Fig.~\ref{fig5} with Figs.~\ref{fig3} and
\ref{fig6}). Correspondingly, the easy-axis anisotropy drives the paraphase at
low temperatures close to long-range order. As can be seen from
Fig.~\ref{fig5}, the validity region of the spin-wave picture, $q \gg
(\xi^{zz})^{-1}$, shrinks with increasing temperature, where predominantly
high-energy magnons may be observed.
\subsection{Spin susceptibility}
The spin anisotropy results in a qualitatively different temperature
dependence of the uniform static susceptibilities $\chi_{0}^{+-}$ and
$\chi_{0}^{zz}$, as can be seen from Fig.~\ref{fig6}. Note that the ED
calculation  of $\chi_{0}^{+-}$ requires a small magnetic field in the $x$
direction so that the $S=1$ data can be obtained only for $N=8$.

The transverse susceptibility $\chi_{0}^{+-}$ (Fig.~\ref{fig6}a) reveals
 a maximum at $T_{m}^{\chi}$, where $T_{m}^{\chi}$ increases with D
(right inset), in very good agreement with the ED results. For small
anisotropies (left inset) a pronounced finite-size effect is observed,
where the theory for $N=8$ agrees well with the ED data. The temperature
dependence of $\chi_{0}^{+-}$ may be explained as follows. The
anisotropy-induced longitudinal SRO (cf.~Fig.~\ref{fig5}) results in a
spin stiffness against the orientation of the transverse spin components
along an external field perpendicular to the $z$ direction. Consequently,
at zero temperature $\chi_{0}^{+-}(0)=2/D$ decreases with increasing
$D$, and at intermediate temperatures $\chi_{0}^{+-}(T)$ exhibits
a maximum.

Considering the longitudinal susceptibility $\chi_{0}^{zz}$
(Fig.~\ref{fig6}b), it shows qualitatively the same behavior as in the
$D=0$ limit; in particular, $\chi_{0}^{zz}$ diverges as $T \to 0$ indicating
the ferromagnetic phase transition. For $T \gtrsim T_{0}$ the Curie law
$\chi_{0}^{+-} = 2 \chi_{0}^{zz} = 4/3T$ holds approximately, where, e.g.,
$T_{0}=2.5 \; (3)$ for $D=0.1 \; (1) $ and $S=1$.
\subsection{Specific heat}
\begin{figure}
\centering
\includegraphics{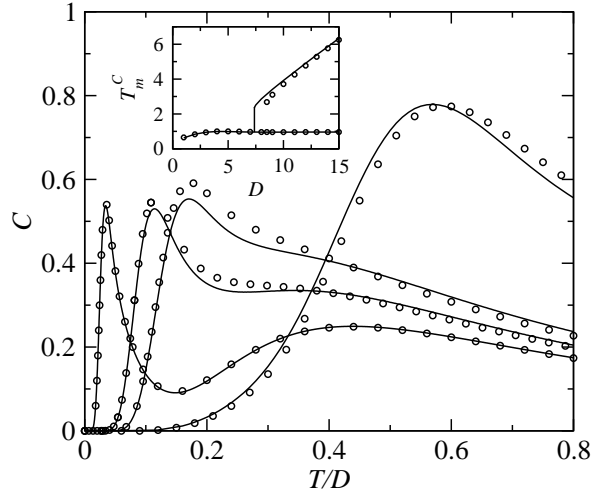}
\caption{Specific heat $C$ for $S=1$ vs $T/D$ at $D=1, 5, 7.4$, and 25,
from right to left, comparing the Green-function (solid) with the ED
($\circ,\; N=12$) results. The inset shows the positions of the maxima
$T_{m}^{C}$ in the temperature dependence of the specific heat as
functions of D.}
\label{fig7}
\end{figure}
\begin{figure}
\centering
\includegraphics{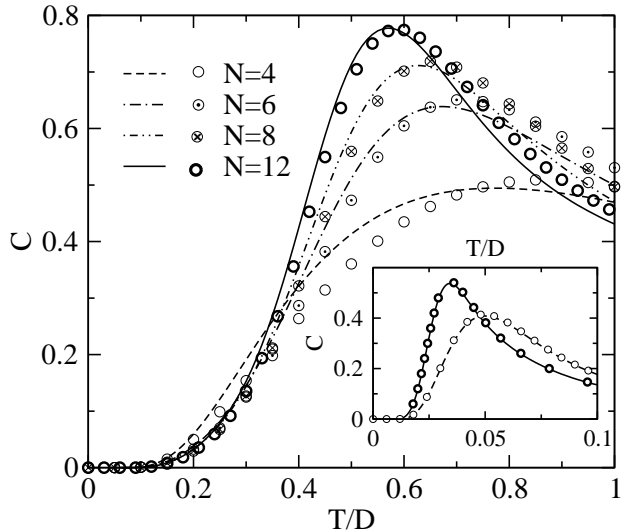}
\caption{ Specific heat for $S=1$ at $D=1$ and $D=25$ (inset: $N=4, 12$) 
and its dependence on the number of spins, where the Green-function results 
(line styles) are compared with the ED data (circles).}
\label{fig8}
\end{figure}
\begin{figure}
\centering
\includegraphics{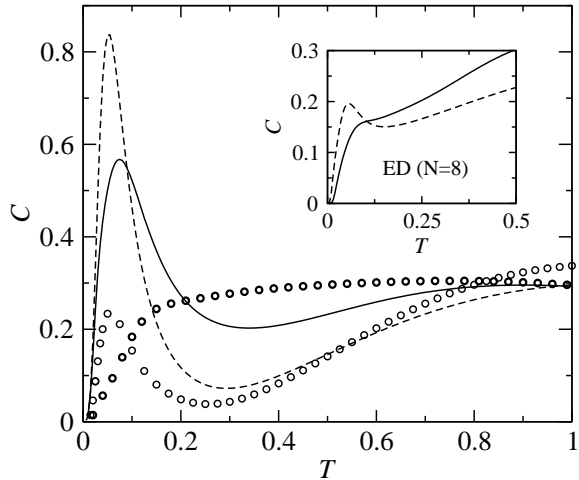}
\caption{Specific heat for $S=1$ at $D=0.1$ and its dependence on the number of
spins, where the ED results for $ N=12$ (\textsf{\textbf{o}}) and $N=4$ ({\sf
o}) are compared with the Green-function theory for $N \rightarrow \infty$
(solid) and $N=4$ (dashed). In the inset, the dependence of the specific heat on
the periodic (solid) versus open (dashed) boundary conditions in the ED
calculations for $N=8$ is demonstrated.}
\label{fig9}
\end{figure}

In Figs.~\ref{fig7} to \ref{fig9} the temperature dependence of the
specific heat for the $S=1$ chain is presented. As the main result,
the ED on $N=12$ chains with periodic boundary conditions yields two
maxima at $T_{m_1}^{C}$ and $T_{m_2}^{C}$, if $D> D_{0}$ with $D_0=7.4$,
and only one maximum at  $T_{m}^{C}$ for
$D<D_{0}$. Let us first consider the specific heat for $D \geqslant 1$
plotted in Fig.~\ref{fig7}. The Green-function results for
$N \to \infty$, agreeing with those for $N=12$ within the accuracy
of drawing, are in a very good agreement with the ED data.
Our results for the maximum
positions nearly agree with those of Bl\"{o}te \cite{BLO75} obtained by
the ED of $N=7$ chains with open boundary conditions and subsequent
extrapolations to $N \to \infty$. For example, for $D=1 \; (5)$ we get
$T_{m}^{C}=0.57 \; (0.85)$, as compared with $T_{m}^{C}=0.5 \; (1.0)$ in
Ref.~\onlinecite{BLO75}; for $D=10 > D_{0}$ we obtain  the maximum
temperatures $T_{m_1}^{C}=0.85$ and $T_{m_2}^{C}=3.93 $ which are slightly
larger than the values  found in Ref.~\onlinecite{BLO75}, $T_{m_1}^{C}=0.79$
and $T_{m_2}^{C}=3.77$. At $D=D_{0}$ the specific heat reveals a plateau
within a small temperature region, $2.4 \lesssim T \lesssim 2.7$ (cf.~
Fig.~\ref{fig7}). Correspondingly, the dependence on $D$ of the maximum
position exhibits a jump at $D_{0}$, as seen in the inset of Fig.~\ref{fig7}.
For $D>D_{0}$, following the reasonings of Ref.~\onlinecite{BLO75} the
upper maximum at $T_{m_2}^{C}$ may be interpreted as Schottky anomaly
due to the $D$ term.

To analyze the finite-size effects on the specific heat for
$D \geqslant 1$ and the accuracy of the Green-function approach in dependence
on the chain length $N$ and on $D$, in Fig.~\ref{fig8} the specific-heat
curves for different values of $N$ with $4 \leqslant N \leqslant 12$ and for
$D=1$ and $D=25$ (see inset) are plotted. As can be seen, the deviation
of the Green-function results from the corresponding ED data decreases
with increasing $N$ and $D$. Comparing the curves for $D=1$ and $D=25$, the
finite-size effects decrease with increasing $D$. Moreover,
they are found to decrease with increasing temperature which is not
shown in Fig.~\ref{fig8}, where, e.g. for $D=25$, only the low-temperature
maximum is depicted (cf.~Fig.~\ref{fig7}).

Considering the specific heat at small anisotropies, the detailed analysis
of our ED calculations for different chain lengths with periodic versus
open boundary conditions reveals considerable finite-size effects, in 
contrast to the case $D \geqslant 1$ discussed above, and
a remarkable dependence of the ED data on the chosen boundary condition.
In this paper we prefer to use periodic boundary conditions, since, due to
the translational symmetry implying equivalent lattice sites,
(i) the finite-size effects are expected to be
less pronounced and (ii) ED calculations for larger systems
($N \leqslant 12$) can be performed, as compared with open boundary
conditions used by Bl\"{o}te\cite{BLO75} for $N \leqslant 7$.
In Fig.~\ref{fig9} we illustrate the finite-size effects and the
influence of boundary conditions for $D=0.1$. The ED data for $N=4$
yield a maximum at $T_{m_1}^{C} \simeq 0.05$ which vanishes for $N=12$.
This may be understood as follows. In finite systems the spin excitations are
gapped, even in the $D=0$ limit, where the finite-size gap $\Delta_{N}$ scales
as $N^{- \alpha}$ with $\alpha >0$. If $D< \Delta_{N}$, a low-temperature
Schottky-type anomaly in the specific heat may appear and vanish for
larger $N$ with $D > \Delta_{N}$. Note that both ED
curves approach each other at $T \gtrsim 5$. In the Green-function theory
for $N=4$ a maximum is also found at the same temperature $T_{m_1}^{C}=0.1$,
but with a too large height. However, for $N \to \infty$ this maximum is only
weakened, but does not disappear. In view of our ED results for $N=12$,
this behavior of the specific heat has to be considered as an artefact
of the Green-function theory for small anisotropies. As can be seen
from the inset of Fig.~\ref{fig9}, the use of open boundary conditions
favors the appearance of a spurious low-temperature maximum in the specific
heat.

In view of our analysis described above, we consider the ED results by
Bl\"{o}te \cite{BLO75} on the specific heat of the
$S=1$ ferromagnetic chain with $D \lesssim 0.25$ as questionable, in
particular, because the extrapolation of the data for small systems
with $N \leqslant 7$ and open boundary conditions was performed.
Our ED results on the specific heat at small anisotropies
qualitatively deviate from the data by  Bl\"{o}te.\cite{BLO75} In
Ref.~\onlinecite{BLO75} two maxima were obtained not only for large values
of $D$ (see above), but also for $D \lesssim 0.25$, where for $D=0.1$
a low-temperature maximum was found at $T_{m_1}^{C}=0.12$. In our ED
data at large enough $N$ such a maximum does not appear.
\subsection{Comparison with experiments}
Finally, let us compare the results of the Green-function theory with some
experiments on Ni complexes \cite{KBD74} and derive predictions for
quantities not yet measured.
\begin{figure}
\centering
\includegraphics{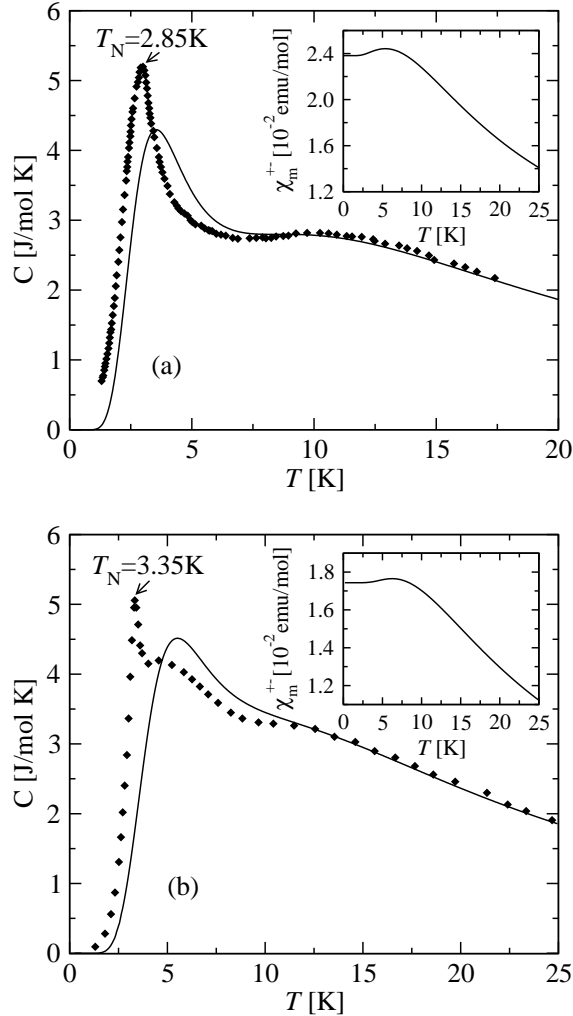}
\caption{Specific heat $C$ of the Ni complexes NiBr$_2 \cdot 2py$ (a)
and NiBr$_2 \cdot 2pz$ (b), where the Green-function theory (solid)
is fit to the experimental data ($ \blacklozenge  $,
Ref.~\onlinecite{KBD74}). The insets show the predicted temperature
dependences of the transverse magnetic susceptibility $\chi_{m}^{+-}$.
}
\label{fig10}
\end{figure}
In Fig.~\ref{fig10} the specific heat of the di-bromo Ni complexes
NiBr$_2$L$_2$ with L=pyrazole ($pz$, N$_2$C$_3$H$_4$) and L=pyridine
($py$, NC$_5$H$_5$) is depicted. Those compounds can be considered as
weakly antiferromagnetically coupled ferromagnetic chains with a large
easy-axis single-ion anisotropy.\cite{KBD74} The small values of the
Ne\'{e}l temperatures $T_{N}$ indicated in Fig.~\ref{fig10} reflect the
pronounced quasi-1D behavior. The anomaly of the specific
heat at $T_{N}$ cannot be described by our theory for a purely 1D
system. For NiBr$_2 \cdot 2 py$ (Fig.~\ref{fig10}a), this anomaly masks
the low-temperature maximum at $T_{m_1}^{C}$. At sufficiently high
temperatures $T>T_{N}$ the systems exhibit 1D behavior,
and the theory may be compared with experiments. For NiBr$_2 \cdot 2py \;(2pz)$
the fit to the specific heat data yields $J=0.4$meV (0.48meV) and $D=3$meV
(2.7meV) so that $D/J=7.5 \;(5.6)$, where the first ratio slightly exceeds
$D_{0}/J$. Note that those values nearly agree with the findings of
Ref.~\onlinecite{KBD74}. Using the fit values for $J$ and $D$ we calculate the
temperature dependence of the transverse magnetic susceptibility
$\chi_{m}^{+-}=4 \mu_{B}^{2} N_A \chi_{0}^{+-}$ ($N_A$ is the Avogadro
constant). The results (see insets of Fig.~\ref{fig10}) show a maximum of
$\chi_{m}^{+-} (T)$ at $T_{m}^{\chi} > T_{N}$, where
\begin{equation}
T_{m}^{\chi}=\left\{ \begin{array}{lll}
5.35 \text{K   } & ; & 2 py \\
6.25 \text{K   } & ; & 2 pz \\
\end{array} \right. ,
\label{tm}
\end{equation}
which should be confirmed experimentally.

\begin{figure}
\centering
\includegraphics{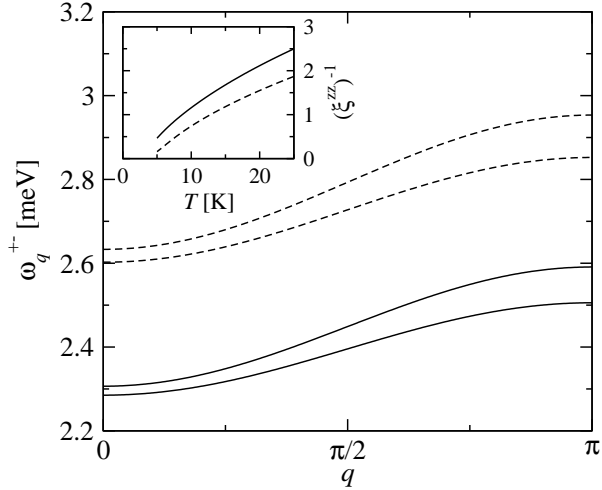}
\caption{Spin-wave spectra $\omega_{q}^{+-}$ for NiBr$_2 \cdot 2py$
(solid) and NiBr$_2 \cdot 2pz$ (dashed) at $T=5$K and 10K, from top
to bottom, and inverse correlation length $ (\xi^{zz})^{-1} $ (inset),
as predicted by the Green-function theory ($q$ and $\xi^{zz}$ are given
in units of the lattice spacing.).}
\label{fig11}
\end{figure}
Furthermore, in Fig.~\ref{fig11} we show the spin-wave spectrum and the
correlation length (inset) calculated for the $J$ and $D$ values given
above. Those results may be verified by neutron scattering experiments
on single crystals. As disscussed in Sec.~A, spin-waves in
the paramagnetic phase may be observed, if $q \gg (\xi^{zz})^{-1}$. For
example, at $T=5$K this condition may be fullfilled for NiBr$_2 \cdot
2 py \; (2 pz)$ with $ (\xi^{zz})^{-1}=0.47 \; (0.16)$. At $T=10$K
we have $(\xi^{zz})^{-1}=1.16 \; (0.74)$ for the $2 py \;
(2 pz)$ complex, so that only Brillouin-zone boundary magnons in NiBr$_2
\cdot 2pz$ may be observable.
\section{SUMMARY}
In this paper we have developed a Green-function theory for $S \geqslant 1$
ferromagnetic Heisenberg chains with an easy-axis on-site anisotropy,
where products of three spin operators are approximated in terms of one
spin operator. Moreover, we have performed exact diagonalizations of chains with
up to $N=12$ sites imposing periodic boundary conditions. To investigate
the spin-wave picture in the paramagnetic phase, we have
calculated the magnon spectrum and the correlation length.
The thermodynamic properties (longitudinal and transverse susceptibilities,
specific heat) at arbitrary temperatures were found to be in good agreement
with the exact results for finite chains. A detailed analysis of the ED data
for the specific heat yields two maxima in the temperature dependence
for $D/J>7.4$, whereas for $D/J<7.4$ only one maximum appears. Our results
at low ratios $D/J$ contradict those of Ref.~\onlinecite{BLO75} obtained
on smaller chains with open boundary conditions. The Green-function theory was
compared with specific heat experiments on di-bromo-pyrazole/pyridine Ni
complexes, and predictions for the spin-wave spectrum, the correlation length,
and the maximum in the temperature dependence of the transverse magnetic
susceptibility were made.

\begin{acknowledgments}
The authors wish to thank K.~Becker and O.~Derzhko for useful discussions.
This work was supported by the Deutsche Forschungsgemeinschaft through the
graduate college "Quantum Field Theory" (I.~J.~J.) and the Projects
RI 615/12-1 and IH 13/7-1. The authors thank J.~Schulenberg for assistance in
ED calculations.
\end{acknowledgments}

\end{document}